\documentclass[conference]{IEEEtran}

\usepackage{url}
\usepackage{cite} 
\usepackage{hyperref}
\usepackage{graphicx}
\usepackage{amsmath,amssymb,amsfonts}
\usepackage{algorithmic}
\usepackage{textcomp}
\usepackage[utf8]{inputenc}

\def\BibTeX{{\rm B\kern-.05em{\sc i\kern-.025em b}\kern-.08em
    T\kern-.1667em\lower.7ex\hbox{E}\kern-.125emX}}

\begin{document}

\title{%
Predicting Startup–VC Fund Matches with Structural Embeddings and Temporal Investment Data
}

\author{
\IEEEauthorblockN{Koutarou Tamura}
\IEEEauthorblockA{ Uzabase, Inc., Japan. \\
Email: koutarou.tamura@uzabase.com, k.tamura.phd@gmail.com}
}

\maketitle

\begin{abstract}
This study proposes a method for predicting startup inclusion, estimating the probability that a venture capital (VC) fund will invest in a given startup. Unlike general recommendation systems, which typically rank multiple candidates, our approach formulates the problem as a binary classification task tailored to each fund–startup pair. Each startup is represented by integrating textual, numerical, and structural features, with Node2Vec capturing network context and multi-head attention enabling feature fusion. Fund investment histories are encoded as LSTM-based sequences of past investees.

Experiments on Japanese startup data demonstrate that the proposed method achieves higher accuracy than a static baseline. The results indicate that incorporating structural features and modeling temporal investment dynamics are effective in capturing fund–startup compatibility.

\end{abstract}


\section{Introduction}
In recent years, investment in startup companies has intensified both domestically and internationally. Accordingly, the importance of making portfolio decisions based on rigorous evaluation of individual firms' growth potential and future prospects has increased. In Japan, notably, the government launched a five-year national startup support initiative led by the Cabinet Office\cite{SU5Y}, resulting in a broader range of industries and business stages becoming eligible for venture funding.

These startup companies are typically supported during their early growth stages through investments from corporate firms and venture capital (VC) funds. The presence or absence of such financial support can exert a significant influence on the trajectory of business development. In recent years, overseas venture capital firms have also been increasing their presence, contributing to a growing volume of investment activity in the Japanese market. In fact, domestic startup funding peaked in 2022 and has since remained at a high level—approximately 800 billion yen annually—with the average amount raised per company continuing to increase \cite{SUFundRep}.

Under such circumstances, fund managers and investment support institutions are required to identify appropriate investment targets from among the growing and diverse pool of emerging startups, and to make informed decisions regarding their inclusion in investment portfolios. In the financial domain, portfolio construction is typically grounded in quantitative assessment and optimization of risk and return. While such evaluations can be conducted based on stock prices in the case of publicly listed companies, unlisted startups generally lack real-time market valuation data. As a result, investment decisions for startups often rely on qualitative information, such as business descriptions and industry classifications. However, these types of information are typically updated infrequently and may fail to capture temporal changes. Consequently, investment decisions are also informed by relational information—such as historical co-investment patterns and shared portfolio holdings with previous funds—indicating that startup evaluation often depends not only on intrinsic attributes, but also on their position within the broader investment network.

To address these challenges, this study proposes a method for estimating the likelihood that a newly emerging startup will be included in a venture capital (VC) fund's portfolio. The approach involves designing both startup and fund embeddings by integrating semantic information—such as firm-specific qualitative and quantitative attributes (e.g., numerical indicators and textual business descriptions)—with structural features derived from inter-firm relationships as captured by historical portfolio co-investment data. This fusion of meaning-based and structure-based features enables a representation that reflects both the intrinsic characteristics of individual firms and their relational positioning within the investment ecosystem.

In this study, we frame startup investment recommendation as an inclusion prediction task. Rather than ranking multiple candidates, we aim to estimate whether a specific candidate startup will be included in a particular VC fund's portfolio, given its historical investments. 
We adopt the term “inclusion prediction” throughout this paper to refer to this fund-specific binary classification task. 

\section{Related Work}

In the context of estimating the likelihood that a startup will be included in a VC fund's portfolio, existing research can be broadly categorized into three major approaches: (1) learning embedding representations of individual companies, (2) modeling company–fund relationships using graph-based representations, and (3) leveraging the historical portfolio composition of VC funds.

\paragraph{Company Representation via Embedding.}
A common approach for representing companies is to learn latent embeddings from textual and numerical information. Company2Vec~\cite{gerling2023company2vec} applies Word2Vec to unstructured corporate website data and generates embeddings that preserve semantic similarity. These have shown improved performance over traditional classification schemes like NACE codes. Other extensions of Company2Vec utilize regulatory filings~\cite{tomokicompany2vec} or LinkedIn-based employee transition networks~\cite{chen2018company2vec}, but these approaches primarily yield static representations.

In contrast, our study combines textual, numerical, and graph-structured inputs to form richer, multifaceted company embeddings that capture both internal characteristics and external structural features.

\paragraph{Graph-Based Modeling of Investment Networks.}
Several studies represent VC investments as bipartite graphs between funds and startups, applying embedding techniques to extract structural features~\cite{lyu2021graph, zhang2015link}. Such networks are interpreted as reflecting strategic and sectoral preferences. Work in economic networks~\cite{KT_IJMPCS, KT_SciRep} has modeled structural flows in supply chains and startup ecosystems~\cite{KT_JSAI2025}, finding that positions within investment graphs signal latent growth potential.

Recent research also models investment dynamics via time-indexed sequences of bipartite graphs, applying graph neural networks (GNNs) with attention mechanisms~\cite{lyu2021graph}. Studies show firms in central positions of startup networks tend to succeed (e.g., IPO or M\&A)~\cite{zhang2015link}, highlighting the predictive power of graph structure.

\paragraph{Combining Network and Temporal Modeling.}
Graph-based structural embeddings, especially those learned via Node2Vec~\cite{grover2016node2vec}, are widely used for capturing structural features. Node2Vec simulates random walks to map nearby nodes to similar vectors, and has been applied to investment graphs to predict link formations. However, such methods often overlook temporal dynamics and ignore node-level attributes.

Our approach addresses these limitations by modeling fund portfolios as time-ordered sequences of company embeddings. Each company embedding combines BERT-based semantic encoding~\cite{BERT,Tohoku_BERT}, numerical indicators, and Node2Vec-based structure. We use LSTM~\cite{hochreiter1997long} to capture the temporal evolution of fund behavior and jointly embed fund and startup entities in a unified framework.

\section{Dataset}

In this study, we conducted experiments using both structural and semantic information on unlisted startup companies in Japan. The dataset was obtained from the startup database provided by Uzabase, Inc.\ through the SPEEDA Startup Information Research platform, reflecting the state as of December 2024.

\subsection{Overview of the Dataset}
Our analysis focuses on investment relationships between domestic startups and investors. 
The dataset includes approximately 24,000 startups and 100,000 investors, with detailed records of investor–startup links for each funding round~\cite{INITIAL_def, INITIAL_num}. 
These investment relationships are primarily extracted from corporate registry records, which provide legally registered information on capital contributions and shareholding status. 
Accordingly, the data mainly captures equity-based investments that are formally recorded, while informal or unregistered financing activities are not included.

To ensure consistency and data reliability, we focus on investments made by corporate entities and VC funds, excluding those made by individuals or entries with unidentified investor types. 
Each investment event records which startup received funding from which investor (i.e., investment fund) at a specific time. 
Although the dataset contains some information on investment amounts, this is often missing and therefore excluded from our analysis.

\subsection{Bipartite Investment Graph Construction}

We construct an undirected bipartite graph where nodes represent startups and investment funds, and edges represent historical investment relationships. Duplicate links and self-loops are removed, and only the first recorded investment between each fund–startup pair is retained.

The final graph includes 12,133 unique startups, 15,189 unique funds, and 63,632 investment edges. The average node degree is approximately 4 to 5, and the degree distribution follows a power-law~\cite{KT_JSAI2025}, indicating a mix of highly active and less active investors.

\subsection{Startup Feature Categories}

Each startup is associated with four types of features:

\begin{itemize}
    \item \textbf{Numerical features:} Company age, number of employees, and cumulative funding amount.
    \item \textbf{Textual features:} Business description text.
    \item \textbf{Categorical features:} Industry classification tags and funding stages.
    \item \textbf{Structural features:} Graph-based information derived from historical investment relationships.
\end{itemize}

These features are preprocessed and later transformed into fixed-dimensional vectors for downstream modeling, as detailed in Section~\ref{sec:company_embedding}.

To ensure completeness and consistency of the startup feature vectors, we apply different imputation strategies tailored to each feature type:
\begin{itemize}
  \item \textbf{Numerical features:} For attributes such as company age or funding amount, if a value is missing for a specific year, it is imputed using the most recent available value from previous years. If all values are missing across time, the geometric mean of the corresponding attribute across all companies is used as the default value.
  \item \textbf{Textual features:} For companies lacking a business description, we insert a placeholder text such as ``no information available,''.
  \item \textbf{Categorical features:} For fields such as geographic location or funding stage (e.g., series A/B), missing values are imputed using the most frequent value among all companies. If an entire categorical field is absent, it is treated as a ``no tag'' entry during encoding.
\end{itemize}
These imputation strategies ensure that all samples maintain structural consistency and remain usable for downstream modeling, regardless of the presence of missing data.

\subsection{Fund Feature Representation}

Each investment fund is represented solely through its historical portfolio within the bipartite investment graph. No explicit fund-level metadata (e.g., strategy or sector focus) is used. The structural relationships in the graph are used to derive fund-level behavioral embeddings in subsequent modeling stages.

\subsection{Transformation of structural features}

To encode inter-company relationships beyond surface-level similarity, we apply Node2Vec to the bipartite investment graph. This generates low-dimensional structural embeddings for each company, reflecting their position within the broader investment network.

\section{Proposed Method}

In this section, we describe our method for predicting whether a given startup will be included in a specific VC fund's portfolio. Our approach consists of four main components: multi-modal company embedding, feature fusion using attention, temporal modeling of fund portfolios, and inclusion prediction based on compatibility scoring.

\subsection{Multi-modal Company Embedding with Attention-based Fusion}
\label{sec:company_embedding}

\begin{figure*}[t]
  \centering
  \includegraphics[width=0.7\linewidth]{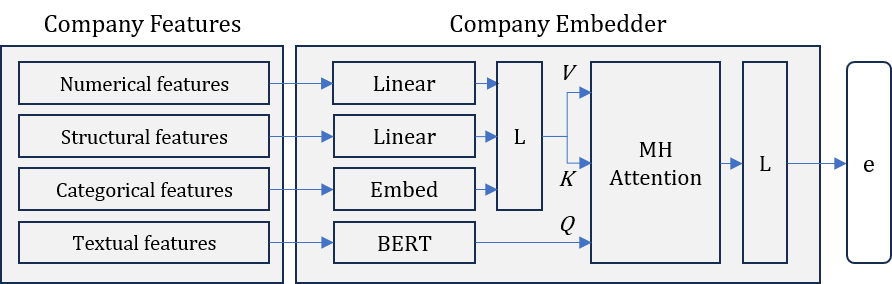}
  \caption{Overview of the startup embedding module. Multi-modal inputs—textual (BERT-based), numerical, categorical, and structural (Node2Vec-based) features—are independently transformed  (with "L" denoting a Linear layer) and then integrated via a multi-head attention mechanism to form the unified startup embedding \( \mathbf{e}_c \).}
  \label{fig:company_embedder}
\end{figure*}

To represent each startup \( c \in C \) as a unified vector \( \mathbf{e}_c \in \mathbb{R}^d \), we integrate four types of features: textual, numerical, categorical, and structural. These modalities collectively encode both firm-specific information and inter-firm relational context, enabling the construction of rich semantic representations suitable for downstream prediction.

\textbf{Textual features:} We use a pre-trained Japanese BERT model~\cite{Tohoku_BERT} to process the business description text of each startup. The [CLS] token embedding from the final encoder layer is extracted as a dense semantic vector representing the textual modality.

\textbf{Numerical features:} Quantitative firm attributes such as company age, number of employees, and cumulative funding amount are normalized and linearly projected into dense embedding spaces. These vectors reflect key size and maturity indicators that may influence investment behavior.

\textbf{Categorical features:} Industry classification tags and funding stages are encoded using either multi-hot vectors or trainable embedding layers, depending on the experimental setting. These features capture domain and lifecycle positioning of the firm.

\textbf{Structural features:} To incorporate relational information, we construct a bipartite graph between startups and VC funds based on historical investment relationships. Node2Vec~\cite{grover2016node2vec} is applied to this graph to generate low-dimensional structural embeddings for each startup. These vectors encode proximity and connectivity within the investment ecosystem, reflecting co-investment patterns and centrality\footnote{We generate 10 random walks of length 20 from each node in the bipartite graph, using Node2Vec parameters $p=1.0$ and $q=0.8$, to train the skip-gram model.}
.

To effectively fuse these heterogeneous feature vectors, we apply a multi-head attention mechanism. The modality-specific embeddings, textual, numerical, categorical, and structural features, are passed through multiple attention heads to compute a weighted combination that reflects the relative importance of each modality. The resulting unified embedding \( \mathbf{e}_c \in \mathbb{R}^{256} \) serves as the final company representation used in subsequent modules.

This embedding process allows the model to adaptively emphasize different modalities depending on the startup's characteristics, such as relying more on structural information for well-connected firms or emphasizing textual content for early-stage startups with limited history.

The overall startup embedding process is summarized in Figure~\ref{fig:company_embedder}, where each modality is transformed independently and then integrated using attention-based fusion.

\subsection{Temporal Modeling of Fund Portfolios}

Each fund \( f \in \mathcal{F} \) is modeled as a time-ordered sequence of the most recent investments. We denote the corresponding input as:

\[
\mathbf{H}_f = (\mathbf{e}_{c_1}, \mathbf{e}_{c_2}, \ldots, \mathbf{e}_{c_T}),
\]

where \( \mathbf{e}_{c_t} \) is the embedding of the \( t \)-th most recent company invested in by fund \( f \), and \( T \leq 15 \). This sequence is passed into an LSTM network to obtain the fund representation:

\[
\mathbf{f}_f = \mathrm{LSTM}(\mathbf{H}_f).
\]

\subsection{Inclusion Prediction via Compatibility Scoring}
\begin{figure*}[t]
  \centering
  \includegraphics[width=0.8\linewidth]{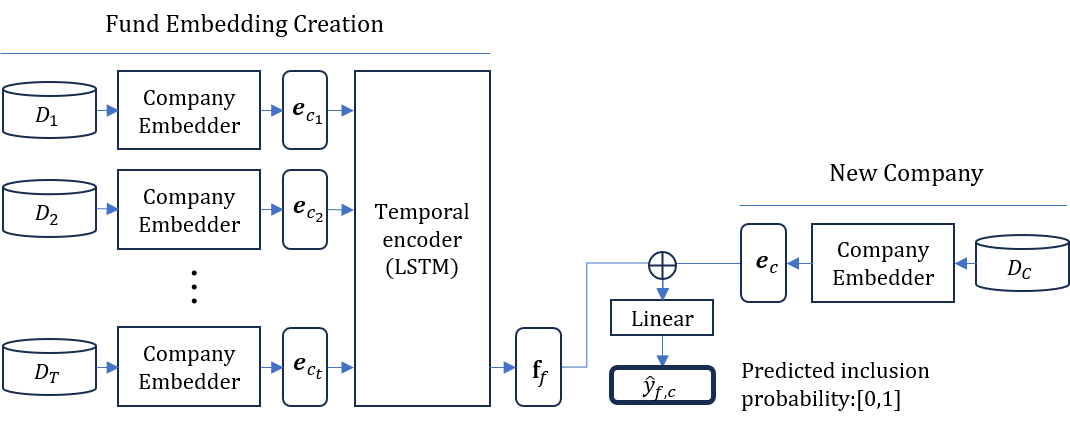}
  \caption{Architecture of the inclusion prediction model. A VC fund is represented as a sequence of past startup embeddings \( \mathbf{e}_c \), encoded via an LSTM to obtain a fund embedding \( \mathbf{f}_f \). This embedding is then combined with a candidate startup embedding to compute the inclusion probability \( \hat{y}_{f,c} \) through a binary classifier.}
  \label{fig:inclusion_model}
\end{figure*}

The overall architecture of our model is shown in Figure~\ref{fig:inclusion_model}. It takes as input a fund embedding \( \mathbf{f}_f \), derived from its past investments via an LSTM, and a candidate company embedding \( \mathbf{e}_c \), and outputs the probability that the fund will invest in that company.

\label{sec:loss}

Given the fund embedding \( \mathbf{f}_f \) and a candidate company embedding \( \mathbf{e}_c \), the model predicts the likelihood that company \( c \) will be included in fund \( f \)'s future portfolio. A binary classifier is trained on pairs \( (\mathbf{f}_f, \mathbf{e}_c) \) using both positive (actual investments) and negative (non-invested) examples.

The prediction output is a scalar probability \( \hat{y}_{f,c} \in [0, 1] \), and the model is trained using the binary cross-entropy loss:

\[
\mathcal{L} = - y \log \hat{y}_{f,c} - (1 - y) \log (1 - \hat{y}_{f,c}),
\]

where \( y \in \{0,1\} \) is the ground-truth label indicating whether the investment occurred.

\vskip\baselineskip
The entire model is trained end-to-end and comprises three core components: 
(1) the company embedding module, which transforms multi-modal startup features into dense embeddings; 
(2) the fund embedding module, which aggregates a variable-length sequence (up to 15) of past investee embeddings using an LSTM; and 
(3) the prediction module, which estimates the inclusion probability based on the compatibility between the fund and candidate embeddings.

All parameters in these modules are jointly optimized via backpropagation, using observed historical investment events as supervision signals. 
This unified training framework enables the model to learn startup representations that are specifically tailored for inclusion prediction, while simultaneously capturing the temporal and behavioral dynamics of individual VC funds.

\section{Experimental Setup}
To evaluate the effectiveness of our proposed inclusion prediction model, we conduct a series of experiments on a real-world Japanese startup investment dataset. This section describes the training procedure, model configurations, and the design of comparative experiments.

To construct the training dataset, we define a reference year \( Y_{\text{cut}} \) that separates historical investment data into context and prediction targets. Specifically, we split the dataset into two subsets: 
(1) past investments \( D^{\text{past}}_{Y_{\text{cut}}} \), used as input to represent fund behavior, and 
(2) future investments \( D^{\text{future}}_{Y_{\text{cut}}} \), used as ground-truth labels for inclusion prediction.

This setup mimics a real-world scenario in which a VC fund makes investment decisions based only on its prior portfolio history, without access to future data. The year \( Y_{\text{cut}} \) is varied across multiple values (2021, 2022, 2023) to construct training examples spanning different time windows. For each fund, we extract up to 15 of the most recent investments prior to \( Y_{\text{cut}} \), and represent them as a time-ordered sequence of company embeddings to compute a fund embedding using an LSTM.

Training examples are formed as fund–company pairs. Positive examples correspond to actual investments made after \( Y_{\text{cut}} \), while negative examples are randomly sampled from companies the fund has never invested in. A 4:6 ratio of positive to negative samples is maintained. Full details of the dataset construction procedure are provided in Appendix~\ref{appendix:dataset}.

Each input pair \( (\mathbf{f}_f, \mathbf{e}_c) \) is passed to a binary classifier that predicts the likelihood of investment \( \hat{y}_{f,c} \in [0,1] \). The model is trained to minimize binary cross-entropy loss, as defined in Section~\ref{sec:loss}.

For companies appearing only in \( D^{\text{future}}_{Y_{\text{cut}}} \)—such as newly founded startups—structural embeddings are unavailable, as they are not included in the Node2Vec training graph. We handle this by imputing missing structural vectors using the mean embedding of all companies in \( D^{\text{past}}_{Y_{\text{cut}}} \). Details are provided in Appendix~\ref{appendix:imputation}.

To evaluate model performance, we randomly split the constructed training dataset into training and validation subsets using a 7:3 ratio. The dataset consists of 88,856 fund–startup investment pairs, each representing a historical inclusion (or non-inclusion) instance. The training set is used to optimize model parameters, while the validation set is used to monitor performance throughout the training process. This split is performed at the level of fund–company pairs to ensure balanced representation across both subsets. Model training is carried out using the Adam optimizer with a learning rate of \(10^{-5}\) and a batch size of 32. Evaluation metrics include precision, recall, and F1-score, which are computed on the validation set to assess the effectiveness of the model. Furthermore, to evaluate generalization capability, we assess model performance on startups that lack structural embeddings by employing different imputation strategies for these missing vectors.

To further assess the contribution of structural features, we conduct an ablation experiment in which the structural embedding component is removed from the model. Specifically, we replace the structural input vector with a fixed zero vector for all companies, while keeping the textual, numerical, and categorical features unchanged. This baseline configuration allows us to isolate the effect of the structural modality, as the model architecture, optimization procedure, and all other inputs remain consistent.

By comparing the performance of this ablated model with the full model that includes Node2Vec-based structural embeddings, we can quantify the added value of relational information captured from the startup–investor network. The performance drop observed in this setting directly reflects the predictive utility of structural features for inclusion prediction.

To test the model's generalization to newly founded startups that lack structural embeddings, we include a setting where such companies—those not present in the original Node2Vec training graph—are assigned the mean embedding vector calculated from observed companies. This enables us to evaluate the model's robustness when relational information is unavailable or incomplete. This configuration is reflected in the third row of Table~\ref{tab:results}.

\section{Results}

We evaluated the performance of the proposed inclusion prediction model under a binary classification setting. 
Fund representations were obtained by integrating sequential startup embeddings using an LSTM-based encoder.

\begin{table}[t]
\centering
\caption{Performance comparison of different model settings.}
\label{tab:results}
\begin{tabular}{l|c|c|c}
\hline
\textbf{Setting} & \textbf{Precision} & \textbf{Recall} & \textbf{F1 Score} \\
\hline
No structural (baseline) & 0.69 & 0.67 & 0.68 \\ \hline
Full structural & 0.76 & 0.74 & 0.75 \\ \hline
Imputed structural (unseen) & 0.69 & 0.73 & 0.71 \\ \hline
\end{tabular}
\end{table}

Table~\ref{tab:results} summarizes the prediction performance under three configurations. 
The first row corresponds to a baseline model that excludes structural information entirely, using only textual, numerical, and categorical features. 
The second row represents the full model, which includes Node2Vec-based structural embeddings. 
The third setting evaluates generalization to startups not present in the structural graph during Node2Vec training; for these unseen startups, structural embeddings are imputed using the average of observed firm embeddings.

As shown in the table, the proposed model achieved an F1 score of 0.75 when full structural information was available, outperforming the baseline by 7 percentage points. 
This improvement confirms that incorporating relational information—captured via graph-based embeddings—significantly enhances the model's ability to predict fund–startup compatibility. 
Structural embeddings appear to encode latent inter-firm connections such as co-investment patterns or shared investors, which are not fully captured by firm-level attributes alone.

Moreover, the model showed stable performance (F1 = 0.71) for startups that lacked structural embeddings, suggesting a degree of robustness. 
We attribute this to the attention-based fusion mechanism, which allows the model to rely more heavily on textual and numerical features when structural information is missing. 
However, some performance degradation remains, highlighting the challenge of imputing structural context and the potential benefit of developing structure-aware embedding methods that generalize better to newly established firms.

\section{Conclusion}

In this study, we proposed a method for inclusion prediction that integrates semantic and structural features of startups and models VC fund behavior using time-ordered sequences of past investments. Our architecture combines multi-modal embeddings via Multihead Attention and captures temporal dynamics through an LSTM-based sequence encoder.

Experiments on Japanese startup data demonstrated a 7-percentage-point improvement in F1 score over the baseline, confirming the effectiveness of incorporating both structural relationships and temporal portfolio transitions. The method also maintained stable performance when predicting for previously unseen startups, indicating its robustness and generalizability.

A key challenge lies in the interpretability and applicability of the model in practice. In the Japanese startup ecosystem, investment decisions are often influenced by relational factors such as investment patterns. Accordingly, capturing these structural contexts contributes not only to prediction accuracy but also to practical insight. As future work, we aim to enhance model transparency via SHAP analysis, attention weight visualization, or rule-based feature extraction that clarifies which factors contribute most to the inclusion prediction.

To evaluate broader applicability, we plan to extend the analysis using international datasets. This will allow us to assess regional differences in investment network structures and test the generalizability of the proposed method across ecosystems.

\bibliographystyle{IEEEtran}

\begin{thebibliography}{99}

\bibitem{SU5Y}
Cabinet Secretariat.
\textit{Startup Development Five-year Plan}.
Apr.~2025.

\bibitem{SUFundRep}
Uzabase Inc.
\textit{Japan Startup Finance}.
Jan.~2025.

\bibitem{gerling2023company2vec}
Christopher Gerling.
Company2Vec – German company embeddings based on corporate websites.
\textit{International Journal of Information Technology \& Decision Making}, 2023.

\bibitem{tomokicompany2vec}
Tomoki Ito, Jose Camacho Collados, Hiroki Sakaji, and Steven Schockaert.
Learning company embeddings from annual reports for fine-grained industry characterization.
In \textit{Proceedings of the 2nd Workshop on Financial Technology and Natural Language Processing (FinNLP 2020)}, 2020.

\bibitem{chen2018company2vec}
Xi Chen, Yiqun Liu, Liang Zhang, and Krishnaram Kenthapadi.
How LinkedIn Economic Graph Bonds Information and Product: Applications in LinkedIn Salary.
In \textit{Proceedings of the 24th ACM SIGKDD International Conference on Knowledge Discovery \& Data Mining},
pp.~120--129, 2018.

\bibitem{lyu2021graph}
Shiwei Lyu, Shuai Ling, Kaihao Guo, Haipeng Zhang, Kunpeng Zhang,
Suting Hong, Qing Ke, and Jinjie Gu.
Graph neural network based VC investment success prediction.
\textit{arXiv preprint} arXiv:2105.11537, 2021.

\bibitem{zhang2015link}
C.~Zhang, E.~Chan, and A.~Abdulhamid.
Link prediction in bipartite venture capital investment networks.
Stanford University CS224W Project Report, 2015.

\bibitem{KT_IJMPCS}
K.~Tamura, W.~Miura, M.~Takayasu, H.~Takayasu, S.~Kitajima, and H.~Goto.
Estimation of flux between interacting nodes on huge inter-firm networks.
\textit{International Journal of Modern Physics: Conference Series}, 16:93--104, 2012.

\bibitem{KT_SciRep}
K.~Tamura, H.~Takayasu, and M.~Takayasu.
Diffusion-localization transition caused by nonlinear transport on complex networks.
\textit{Scientific Reports}, 8(5517), 2018.

\bibitem{KT_JSAI2025}
Koutarou Tamura.
Network analysis of investment relationships in startups.
In \textit{Proceedings of the 39th Annual Conference of the Japanese Society for Artificial Intelligence (JSAI2025)},
2025. (In Japanese)

\bibitem{grover2016node2vec}
Aditya Grover and Jure Leskovec.
node2vec: Scalable feature learning for networks.
In \textit{Proceedings of the 22nd ACM SIGKDD International Conference on Knowledge Discovery and Data Mining},
pp.~855--864, 2016.

\bibitem{BERT}
J.~Devlin, M.~Chang, K.~Lee, and K.~Toutanova.
BERT: Pre-training of deep bidirectional transformers for language understanding.
\textit{arXiv:1810.04805}, 2019.

\bibitem{Tohoku_BERT}
Tohoku University.
\textit{cl-tohoku/bert-base-japanese-v3: Pretrained BERT model for Japanese}.
Available at: \url{https://huggingface.co/cl-tohoku/bert-base-japanese-v3}.
Accessed Jan.~2023.

\bibitem{hochreiter1997long}
Sepp Hochreiter and Jürgen Schmidhuber.
Long short-term memory.
\textit{Neural Computation}, 9(8):1735--1780, 1997.

\bibitem{INITIAL_def}
Uzabase Inc.
Definition of Startups.
Available at: \url{https://help.initial.inc/ja/articles/4570807}.
Accessed Feb.~2025.

\bibitem{INITIAL_num}
Uzabase Inc.
Overview of Data Collection.
Available at: \url{https://help.initial.inc/ja/articles/4607633}.
Accessed Feb.~2025.

\end{thebibliography}

\appendices
\section{Dataset Construction Procedure}
\label{appendix:dataset}

Let \( D = \{(f, c, t)\} \) be the historical investment dataset, where \( f \in \mathcal{F} \) is a VC fund, \( c \in \mathcal{C} \) is a startup, and \( t \in \mathbb{R} \) is the time of investment.

For each cutoff year \( Y_{\text{cut}} \), we define:

\begin{itemize}
  \item \textbf{Past investments (training context):}
  \[
  D^{\text{past}}_{Y_{\text{cut}}} = \{(f, c, t) \in D \mid t \leq Y_{\text{cut}}\}
  \]

  \item \textbf{Future investments (prediction targets):}
  \[
  D^{\text{future}}_{Y_{\text{cut}}} = \{(f, c, t) \in D \mid t > Y_{\text{cut}}\}
  \]
\end{itemize}

\subsubsection*{Positive Examples}  
We define the set of positive examples as:
\[
\mathcal{D}^{\text{pos}}_{Y_{\text{cut}}} = \left\{(\mathbf{f}_f, \mathbf{e}_c, 1) \mid (f, c, t) \in D^{\text{future}}_{Y_{\text{cut}}} \right\}
\]

\subsubsection*{Negative Examples}  
For each fund \( f \), let \( \mathcal{C}_f = \{ c \mid \exists t,\; (f, c, t) \in D^{\text{past}}_{Y_{\text{cut}}} \} \) be the set of companies previously invested in. We define negative examples as:
\[
\mathcal{D}^{\text{neg}}_{Y_{\text{cut}}} = \left\{(\mathbf{f}_f, \mathbf{e}_c, 0) \mid f \in \mathcal{F},\; c \in \mathcal{C} \setminus \mathcal{C}_f \right\}
\]
where \( c \in \mathcal{C} \setminus \mathcal{C}_f \) indicates that startup \( c \) has not been previously invested in by fund \( f \).  
We randomly sample negative pairs to maintain a 4:6 ratio with the positive set.

\subsubsection*{Merged Training Set}  
For each year \( Y_{\text{cut}} \in \{2021, 2022, 2023\} \), we construct:
\[
\mathcal{D}_{Y_{\text{cut}}}^{\text{train}} = \mathcal{D}^{\text{pos}}_{Y_{\text{cut}}} \cup \mathcal{D}^{\text{neg}}_{Y_{\text{cut}}}
\]
Finally, to build a comprehensive training dataset covering multiple temporal contexts, we aggregate training examples from all cutoff years:
\[
\mathcal{D}^{\text{train}} = \bigcup_{Y_{\text{cut}} \in \{2021, 2022, 2023\}} \mathcal{D}_{Y_{\text{cut}}}^{\text{train}}
\]

\section{Handling Missing Structural Embeddings}
\label{appendix:imputation}
Companies that appear only in \( D^{\text{future}}_{Y_{\text{cut}}} \) (i.e., not present in the investment history prior to the cutoff year) do not appear in the Node2Vec training graph. Consequently, structural embeddings cannot be directly obtained for these companies.

To address this issue, we impute the structural embedding for such companies using the mean vector computed across all company embeddings in \( D^{\text{past}}_{Y_{\text{cut}}} \). Formally, for any company \( c \in D^{\text{future}}_{Y_{\text{cut}}} \setminus D^{\text{past}}_{Y_{\text{cut}}} \), i.e., startups that appear only after the cutoff year and were not observed in the structural graph, we define its structural component as:

\[
\mathbf{s}_c = \frac{1}{|\mathcal{C}_{\text{past}}|} \sum_{c' \in \mathcal{C}_{\text{past}}} \mathbf{s}_{c'} \]
where,  $ \mathcal{C}_{\text{past}} = \{ c' \mid (f, c', t) \in D^{\text{past}}_{Y_{\text{cut}}} $

\end{document}